\RequirePackage{fix-cm}
\documentclass[twocolumn]{svjour3}          % twocolumn
\smartqed  % flush right qed marks, e.g. at end of proof
\usepackage{graphicx}
\newcommand{\AU}{au}
\newcommand{\MN}{Mon.\ Not.\ R. Astron.\ Soc.}
\newcommand{\AAp}{Astron.\ Astrophys.}
\newcommand{\AJ}{Astron.\ J.}
\newcommand{\ApJ}{Astrophys.\ J.}
\newcommand{\CelMech}{Celest.\ Mech.\ Dyn.\ Astron.}
\newcommand{\EMP}{Earth Moon Plan.}
\newcommand{\Nat}{Nature}
\newcommand{\Sci}{Science}

\newcommand{\ltsimeq}{\raisebox{-0.6ex}{$\,\stackrel
                      {\raisebox{-.2ex}{$\textstyle <$}}{\sim}\,$}}
\newcommand{\gtsimeq}{\raisebox{-0.6ex}{$\,\stackrel
                      {\raisebox{-.2ex}{$\textstyle >$}}{\sim}\,$}}

\newcommand{\mc}{\multicolumn{2}{c}}

\journalname{Earth, Moon, and Planets}
\begin{document}

\title{A model for the common origin of Jupiter family and Halley type
       comets}%\thanks{Grants or other notes


\titlerunning{Common origin of Jupiter family and Halley type comets}
              % if too long for running head

\author{V.V. Emel'yanenko \and
        D.J. Asher        \and
        M.E. Bailey %etc.
}

\authorrunning{Emel'yanenko et al.} % if too long for running head

\institute{V.V. Emel'yanenko  \at
              Institute of Astronomy RAS, 48 Pyatnitskaya, Moscow, 119017,
              Russia \\
              \email{vvemel@inasan.ru}           %  \\
%            \emph{Present address:} of F. Author  %  if needed
            \and
           D.J. Asher, M.E. Bailey  \at
              Armagh Observatory, College Hill, Armagh, BT61 9DG,
              United Kingdom
}

\date{Received: date / Accepted: date}

\maketitle

\begin{abstract}
A numerical simulation of the Oort cloud is used to explain the observed
orbital distributions and numbers of Jupiter-family and Halley-type
short-period comets.  Comets are given initial orbits with
perihelion distances between 5 and 36 \AU, and evolve
under planetary, stellar and Galactic perturbations for 4.5 Gyr.  This
process leads to the formation of an Oort cloud (which we define as the
region of semimajor axes $a>1000$ \AU), and to a flux of cometary bodies
from the Oort cloud returning to the planetary region at the present
epoch.  The results are consistent with the dynamical characteristics of
short-period comets and other observed cometary populations: the
near-parabolic flux, Centaurs, and high-eccentricity trans-Neptunian
objects.  To achieve this consistency with observations, the model
requires that the number of comets versus initial perihelion distance is
concentrated towards the outer planetary region.  Moreover, the mean
physical lifetime of observable comets in the inner planetary region
($q<2.5$ \AU) at the present epoch should be an increasing function of
the comets' initial perihelion distances.  Virtually all observed
Halley-type comets and nearly half of observed Jupiter-family comets
come from the Oort cloud, and initially (4.5 Gyr ago) from orbits
concentrated near the outer planetary region.  Comets that have been in
the Oort cloud also return to the Centaur ($5\!<\!q\!<\!28$ \AU,
$a<1000$ \AU) and near-Neptune high-eccentricity regions.  Such objects
with perihelia near Neptune are hard to discover, but Centaurs with
characteristics predicted by the model (e.g.\ large semimajor axes,
above 60 \AU, or high inclinations, above $40^\circ$) are increasingly
being found by observers.  The model provides a unified picture for the
origin of Jupiter-family and Halley-type comets.  It predicts that the
mean physical lifetime of all comets in the region $q<1.5$ \AU\ is less
than $\sim$200 revolutions.
\keywords{Comets \and Oort cloud \and Centaurs \and Solar system formation
          \and celestial mechanics}
\end{abstract}

\section{Introduction}

\subsection{Origin of short-period comets}

Explaining the origin of short-period (SP) comets (periods $P<200$ yr) is a
long-standing problem. The main difficulty lies in the differences and
apparent inconsistency between the respective numbers and orbital
distributions of Jupiter-family (JF) and Halley-type (HT) comets.
These we classify using the Tisserand parameter $T$ with respect to Jupiter
(Carusi et al.\ 1987), JF comets having $T>2$ (and $P$ usually below 20
years), HT comets having $T<2$ (and $P$ usually between 20 and 200
years). When SP comets are classified this way the number of observed HT
comets is found to be less than, or at most comparable to, the number of
observed JF comets (see Section \ref{obs-sp} below).  However, most
dynamical theories of their origin from the near-parabolic flux predict
a far greater proportion of HT comets (Emel'yanenko and Bailey 1998),
with the overall number of observed JF comets conversely being much too
large relative to the calculated number (Joss 1973; Delsemme 1973). This
discrepancy is associated with the well-known fading problem for
long-period comets originating in the Oort cloud and has, at least in part,
led to the idea that the two classes of SP comet may have
different physical structures and different proximate sources in the
present Solar system.

Thus, although there have been many advances in
understanding the diverse populations of small bodies in the Solar system,
neither a single source dominated by trans-Neptunian objects nor one
dominated by the traditional Oort cloud near-parabolic flux at small
perihelion distances seems capable of explaining the entire distribution of
orbital elements of SP comets. In particular the observed JF comet
inclination distribution was recognized to have too many comets at low $i$
relative to the calculated distribution (Duncan et al.\ 1988; Quinn et
al.\ 1990).

For these reasons, the majority of authors nowadays consider JF and HT
comets to be physically as well as dynamically distinct classes, presumably
formed in separate regions of the early Solar system and having different
dynamical and physical evolutionary histories. Under this viewpoint, JF
comets are often regarded as originating largely in the proto-planetary
disc beyond Neptune, for example in or close to the Edgeworth-Kuiper
belt (EKB). The idea that JF comets might originate in a primordial disc
or `belt' of comets located near or beyond the orbit of Neptune was
investigated by a number of authors (e.g.\ Fern\'andez 1980, 1982;
Duncan et al.\ 1988; Torbett 1989; Torbett and Smoluchowski 1990; Quinn
et al.\ 1990). The discovery of 1992 QB$_1$ (Jewitt and Luu 1993) and
of further Edgeworth-Kuiper objects played a pivotal role in theories of
the origin of SP comets, and important advances building on this
evidence were made in particular by Duncan et al.\ (1995), Duncan and
Levison (1997), Levison and Duncan (1997), and Levison et al.\ (2001). A
key point (Duncan and Levison 1997) was recognition of the potentially
important role played by the `scattered' disc, introduced by Torbett
(1989) and detected a few years later (Luu et al.\ 1997), in which it
appears that the scattered disc of primordial objects originally formed
in the region of the major planets is the principal source of observed
JF comets, rather than the EKB. Under the viewpoint of
distinct JF and HT classes, HT comets are regarded
as objects captured from the Oort cloud (Levison et al.\ 2001), a
structure that would have been produced inevitably as a by-product of
planetary, stellar and other perturbations acting on planetesimals or
cometary nuclei originally formed by accretion within the planetary region
of the proto-planetary disc.

However, a rather unsatisfactory feature of this general picture is the
assumption that HT comets coming from the Oort cloud must disintegrate very
quickly in order to explain the small number of objects observed
(Emel'yanenko and Bailey 1998; Levison et al.\ 2001). The
number of observed inert HT `asteroids' is also very small, and it seems as
if the disintegration of a kilometre-size comet nucleus, into presumably an
initial trail of much smaller boulder-size objects and then finally dust,
must proceed fairly rapidly and lead to eventual extinction of the original
comet. On the other hand, dynamical theories
appear to require that a high proportion of the JF comet source flux should
survive dynamical transfer into the inner Solar system to become active JF
comets and that these JF comets should survive for $\sim$10$^3$
revolutions in the inner Solar system.
This difference in the physico-dynamical evolution of the two types
of objects is the fading problem for SP comets.

It is probably not unreasonable to assume that comets that formed in
different parts of the proto-planetary disc have different physical
properties and therefore different lifetimes in the observable region, and
it appears that this idea has become very deeply rooted.  What is missing,
however, is direct observational evidence to support the idea of two
qualitatively distinct types of SP comet, correlating
with dynamical class.
Thus, present theories of the origin of SP comets rely on a poorly
understood fading hypothesis to accommodate the observations, and there is
no satisfactory physico-dynamical explanation as to why two very different
types of SP comet should exist and yet appear observationally almost
indistinguishable.  Indeed, although comets show a very
diverse range of properties, covering a very broad range of sizes,
densities, dust-to-gas ratios and so on, there is as yet no compelling
observational evidence for the expected bimodality of physical
characteristics corresponding to HT versus JF dynamical class
(Lamy et al.\ 2004).

In this work, whilst recognizing that comets may have
different physical properties depending for example on their sizes or where
they might have formed in the proto-planetary disc, we present a model
for the common origin and evolution -- from the Oort cloud --
of the majority of comets in the
Solar system.

\subsection{Role of the Oort cloud}

We define the Oort cloud as the region containing objects with
semimajor axes $a>10^3$ \AU\ (i.e., objects from the Oort cloud have
at some point during their evolution reached $a>10^3$ \AU).  This
definition is consistent with those used by other authors; e.g.\
Wiegert and Tremaine (1999) and Rickman et al.\ (2008) used similar values
of semimajor axis, i.e. $a\simeq \mbox{1--3}\times 10^3$ \AU, to define the
inner boundary of the Oort cloud.  Dones et al.\
(2004) introduced a further restriction, namely that the maximum value of
perihelion distance
$q$ during an object's orbital evolution should exceed 45 \AU\ for it to be
counted as an `Oort cloud' object.  However, objects with
$a \gtsimeq 10^3$ \AU\ spend nearly all their time at large heliocentric
distances, whatever their value of $q$, and therefore in the Oort cloud.

In this paper we have chosen to define Oort-cloud objects
solely according to $a$ because the
influence of stellar and Galactic perturbations is determined mainly by $a$
for near-parabolic orbits. It has been shown (e.g.\ Emel'yanenko
2005) that the dynamical pathways by which objects with $a>10^3$~\AU\
reach the planetary region are different from those of typical
trans-Neptunian objects (TNOs). While the evolution of TNOs is
largely determined only by planetary perturbations, stellar and Galactic
perturbations play a more substantial role in the process that drives the
perihelia of objects with $a>10^3$ \AU\ towards and through
the planetary region,
regardless of their previous $q$.

In the present paper (Section \ref{integ} onwards)
we numerically integrate a much larger number of
objects than in the Oort cloud model of
Emel'yanenko et al.\ (2007), in particular to obtain
statistically significant numbers of SP comets captured from the Oort
cloud and allow a comparison of the model SP numbers and orbital
distributions with the corresponding distributions of observed
HT and JF comets.  First (Section \ref{obs-com}), in order that our
model parameters can be constrained by observations, we assess the known
characteristics of the various populations of cometary bodies.

\section{Principal features of observed cometary populations}
\label{obs-com}

\subsection{Short-period comets}
\label{obs-sp}

We took data from the MPC (Minor Planet Center) and JPL (Jet Propulsion
Laboratory) lists of discovered comets with $P<200$ yr and
$q<1.5$ \AU\ near the present epoch.
The completeness level in the discovery of
SP comets is slightly uncertain, especially
for HT comets and when results are extrapolated to fainter magnitudes and
larger perihelion distances.  However, many discussions (e.g.\ Fern\'andez
et al.\ 1999; Levison et al.\ 2001) have indicated a relatively high degree
of completeness in the observed sample of active comets at small perihelion
distances ($q<1.5$ \AU).
This level of completeness is supported too by studies of long-period
comets, essentially none of which have been missed at $q<1.3$ \AU\ since
1985 (Fern\'andez and Sosa 2012).

We excluded SOHO comets because these
have rather uncertain physical and dynamical characteristics; this only
affects the distribution near very small $q$, a region that we do not study
here.  We also excluded multiple-apparition comets that have not been
observed for a number of revolutions and are now treated as dead or inert.
For split comets we took only the orbit of the main nucleus. In the end we
obtained a list of 103 observed objects that we regard as representing the
present-day set of active SP comets with $q<1.5$ \AU.  Of these, 75 have
$T>2$ (JF comets) and 28 have $T<2$ (HT comets).

\begin{figure}
\includegraphics[width=\columnwidth]{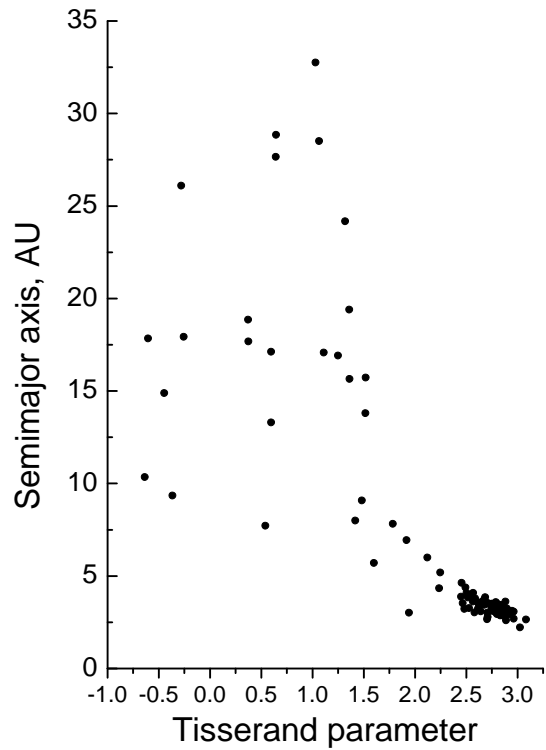}
\caption{The distribution of $T$ and $a$ for
observed short-period comets with $q<1.5$ \AU.
}
\label{atql1.5}
\end{figure}

\begin{figure}
\includegraphics[width=\columnwidth]{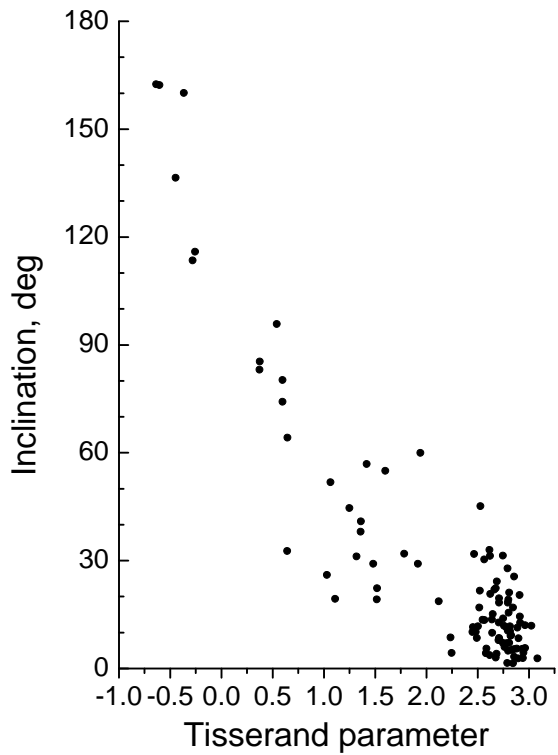}
\caption{The distribution of $T$ and $i$ for
observed short-period comets with $q<1.5$ \AU.
}
\label{itql1.5}
\end{figure}

Figures \ref{atql1.5} and \ref{itql1.5} present orbital element distributions
of these observed objects.  The inclinations (Figure \ref{itql1.5}) show JF
comets ($T>2$) are concentrated close to the ecliptic and prograde HT comets
outnumber retrograde ones (Fern\'andez and Gallardo 1994; Levison et al.\
2001).

Additionally the intrinsic numbers of JF and HT comets are a key constraint
for our model.  Fern\'andez et al.\ (1999) found that about a hundred active
JF comets should currently exist in the region $q<1.5$ \AU, down to nuclear
radius $R$$\sim$0.7 km.  The number appears to drop very rapidly for smaller
bodies (Fern\'andez et al.\ 1999; Snodgrass et al.\ 2011).
This estimate could be
modified to take account of more recent comet discoveries (cf.\ Section 2.1
of Di Sisto et al.\ 2009) but the result would not be significantly changed.

For HT comets, their longer average orbital periods constitute the
principal bias against their discovery relative to JF comets.  That is,
although we expect that most active comets passing perihelion with
sufficiently small $q$ will be found at the current level of
observational searches, many HT comets have not yet returned to
perihelion during the last few decades when searches have been at such
levels.  In this way, taking account of the HT period distribution,
Levison et al.\ (2001), from 22 observed HT comets with $q<1.3\,$\AU,
estimated a population of 57 active HT comets ($q<1.3$).  This result
can be extrapolated to about a hundred objects with $q<1.5$ \AU.
In a later paper (Levison et al.\ 2006) the observed number has only
increased to 24, suggesting the estimate is reliable.

We conclude that there are
roughly a hundred active JF comets and a comparable number, i.e.\
approximately a hundred, of active HT comets to be explained in the region
$q<1.5$ \AU\ at times near the present epoch.
Certainly the number of already known active JF comets shows that their
intrinsic number cannot be much below a hundred, while the intrinsic HT
number cannot be much above a hundred without implausibly many bright
comets being missed by observational searches.

\subsection{Near-parabolic flux}
\label{nr-parab}

The flux of dynamically new comets from the Oort cloud is a fundamental
parameter underpinning all dynamical models of the small-body populations
in the Solar system, including the estimates in this paper.
There are uncertainties in the
frequency, $\nu_{\mbox{\scriptsize new}}$, of comets with $a>10^4$ \AU\ passing
perihelion per \AU\ in $q$ per year, but $\nu_{\mbox{\scriptsize new}}$ is usually
estimated to lie in the range 2 to 4 for present-day comets in near-Earth
space (Bailey and Stagg 1988; Fern\'andez and Gallardo 1999; Wiegert and
Tremaine 1999; Francis 2005).  For quantitative estimates in this paper we
adopt $\nu_{\mbox{\scriptsize new}}=2.5$, within the observable region
$q<1.5$ \AU.

Francis (2005) undertook a detailed discussion of the objects
that the LINEAR survey should discover for a given intrinsic cometary
population. Considering also the question of the cometary absolute
magnitude distribution, he found that very faint (on average presumably
smaller) comets are only slightly more abundant than somewhat brighter
(presumably larger) ones.  Thus statements about cometary numbers, while
evidently depending in detail on the adopted absolute-magnitude cutoff, are
not strongly dependent on the precise value of that cutoff.  In order to
fix ideas, our adopted value $\nu_{\mbox{\scriptsize new}} = 2.5$ comets
with $a>10^4$ \AU\ passing perihelion per \AU\ in $q$ per year may be assumed
to apply to comets with total visual
absolute magnitudes $H_{10} \ltsimeq 11$.  The quantity $H_{10}$ is the
magnitude normalized to 1 \AU\ from Earth and Sun (e.g.\ Everhart 1967).
The inclusion of fainter comets (e.g.\ extrapolating results from
$H_{10}$ = 11 to $H_{10} = 16$) makes very little practical
difference to our results (Francis 2005; Sosa and Fern\'andez 2011), although
the calibration factor,
$\nu_{\mbox{\scriptsize new}}$, would of course increase.
The relative lack of very small (diameters $d \ltsimeq 0.5$ km)
comets (Fern\'andez and Sosa 2012) suggests that
the physical response of the smallest dynamically `new' comets from the
Oort cloud to the thermal shock of their first passage at small perihelion
provides a clue to the underlying rapid fading of new comets from the Oort
cloud, necessary to explain the detailed shape of the observed
$1/a$-distribution (cf.\ Bailey 1984).

\begin{table}
\begin{center}

\caption{Centaurs (objects with $5<q<28$ \AU\ and $a<1000$\ \AU,
excluding a few resonant trans-Neptunian objects and Trojans) that have a
probable source in the Oort cloud. The majority of such objects have $a>60$
\AU\ and after observational debiasing would be extremely numerous.
Centaurs with $a<60$ \AU\ are listed if $i>40^\circ$. Only Centaurs with an
observational arc larger than 100 days (asteroid orbits from
MPC) and comets of orbital Classes 1 and 2 (Marsden and Williams
2008) are included. A unified classification scheme for Centaurs was
proposed by Horner et al.\ (2003).}
\label{cco}

\begin{tabular}{@{}r@{\,\,}lr@{}lrr@{}}
\hline
&& \multicolumn{2}{c}{$a$}
 & \multicolumn{1}{c}{$q$}
 & \multicolumn{1}{c@{}}{$i$}\\
&& \multicolumn{2}{c}{\AU}
 & \multicolumn{1}{c}{\AU}
 & \multicolumn{1}{c@{}}{deg}\\[5pt]
 (29981) & 1999 TD$_{10}$   &  99&.4 & 12.3 &   6 \\
 (87269) & 2000 OO$_{67}$   & 653&   & 20.8 &  20 \\
(127546) & 2002 XU$_{93}$   &  66&.8 & 21.0 &  78 \\
         & 2003 FH$_{129}$  &  71&.3 & 27.6 &  19 \\
 (65489) & 2003 FX$_{128}$  & 100&   & 17.8 &  22 \\
         & 2004 VH$_{131}$  &  60&.8 & 22.3 &  12 \\
         & 2005 VD          &   6&.7 &  5.0 & 173 \\
(308933) & 2006 SQ$_{372}$  & 906&   & 24.2 &  19 \\
         & 2007 JK$_{43}$   &  46&.1 & 23.6 &  45 \\
         & 2007 UM$_{126}$  &  12&.9 &  8.5 &  42 \\
         & 2008 KV$_{42}$   &  41&.7 & 21.2 & 104 \\
(315898) & 2008 QD$_{4}$    &   8&.4 &  5.4 &  42 \\
         & 2008 YB$_{3}$    &  11&.7 &  6.5 & 105 \\
         & 2009 MS$_{9}$    & 386&   & 11.0 &  68 \\
         & 2009 YD$_{7}$    & 129&   & 13.4 &  31 \\
         & 2010 BK$_{118}$  & 447&   &  6.1 & 144 \\
         & 2010 JJ$_{124}$  &  82&.9 & 23.6 &  38 \\
         & 2010 NV$_{1}$    & 294&   &  9.4 & 141 \\
         & 2010 WG$_{9}$    &  53&.8 & 18.8 &  70 \\
         & C/1984 U1        & 646&   &  5.5 & 179 \\
         & C/1998 M6        & 972&   &  6.0 &  92 \\
         & C/1999 K2        & 145&   &  5.3 &  82 \\
         & C/2001 Q1        & 176&   &  5.8 &  67 \\
         & C/2002 K2        & 561&   &  5.2 & 131 \\
         & C/2002 P1        & 497&   &  6.5 &  35 \\
         & C/2002 VQ$_{94}$ & 189&   &  6.8 &  71 \\
         & C/2003 J1        & 514&   &  5.1 &  98 \\
         & C/2005 R4        & 914&   &  5.2 & 164 \\
         & C/2007 D3        & 765&   &  5.2 &  46 \\
         & C/2007 K1        & 425&   &  9.2 & 108 \\
\hline
\end{tabular}

\end{center}
\end{table}

\subsection{Centaurs}

Centaurs are an intermediate cometary population (including
active comets and inactive apparent asteroids), some of them being
en-route from the outer Solar system to near-Earth space and the
SP comet region.  As a transition population the Centaurs must be
replenished from a more distant source, presumably located either in the
trans-Neptunian region or the Oort cloud, and they play a pivotal role in
constraining theories of the origin of SP comets.  

There is however no abiding consensus on the exact definition of a
Centaur.  Many authors (e.g.\ Stern and Campins 1996; Gladman 2002;
Gladman et al.\ 2008; Jewitt 2009) adopt the criterion
that a Centaur should orbit largely in the region of the outer planets.
This has often been taken to mean $a \ltsimeq 30$~\AU, i.e.\ less than
the semimajor axis of Neptune.  In contrast, following Emel'yanenko et
al.\ (2007), we define Centaurs as small bodies moving in heliocentric
orbits with $5<q<28$ \AU\ and $a<1000$ \AU\ (with any value of $i$),
excluding a few resonant trans-Neptunian objects and Trojans.  Thus,
many objects that we call Centaurs (cf.\ Horner et al.\ 2003,
2004a,b) would be classified by some other authors as scattered-disc
objects.

The condition $q<28$ \AU\ separates Centaurs from the NNHE
region described in Section \ref{tno}.
Our Centaur definition reflects the fact that this entire region of
orbital element phase space ($a<1000$ \AU\ and any $i$) constitutes a
transition region of dynamically short-lived orbits in which population
numbers and orbit distributions provide vital evidence about the outer
Solar system source regions.  So whereas a significant number of
Centaurs are produced by dynamical evolution from the Kuiper
belt or the trans-Neptunian region
(e.g.\ Tiscareno and Malhotra 2003; Volk and
Malhotra 2008), we emphasize that using a similar definition of a Centaur to
that used in this paper, Emel'yanenko et al.\ (2005) showed the
debiased distribution of observed Centaurs contradicts the idea that
Centaurs {\em primarily\/} originate from a flattened disc-like
population.  They inferred instead that the Oort cloud produces
$\sim$90\% of Centaurs, specifically well over 90\% of
Centaurs that have $a>60$ \AU\ (which themselves constitute 90\%
of the Centaur population after observational debiasing) and $\sim$50\%
of Centaurs with $a<60$~\AU.  Of these $a<60$ Centaurs, the
Oort cloud contributes especially to those with $i>40^\circ$.

Observational evidence for Centaurs with these orbital characteristics
is growing (Table \ref{cco}),
consistent with predictions (Emel'yanenko 2005;
Emel'yanenko et al.\ 2005) that a significant number of Centaurs have a
proximate source in the Oort cloud.  Emel'yanenko et al.\ (2005)
concluded that there were two separate but overlapping dynamical classes
of Centaurs, one originating in the Oort cloud and the other from the
observed near-Neptune high-eccentricity region, each source region producing
$\sim$50\% of Centaurs with $a \ltsimeq 60$ \AU\ and $\sim$50\%
of JF comets.  A bimodal colour distribution is observed in Centaurs
(Peixinho et al.\ 2003).  The only presently apparent difference in the
two groups' orbital properties is that red Centaurs tend to have lower
$i$ (Tegler et al.\ 2008), while Peixinho et al.\ (2012) instead find
that the bimodality is only pronounced in smaller objects.
A dynamical evolution study suggests red
Centaurs have spent less time at small $q$ (Melita and Licandro 2012).

\subsection{Trans-Neptunian objects}
\label{tno}

As with Centaurs the nomenclature is not universal.  For example (Gladman
et al.\ 2008) in some classification schemes the term `Kuiper belt' can mean
the union of the `classical' Kuiper belt, the scattered disc, the
`extended' (or detached) scattered disc and resonant objects exterior to
the Neptune Trojans, the whole region sometimes being described simply as
the trans-Neptunian region.

We define the trans-Neptunian region as the part of the Solar system in
the vicinity of and beyond Neptune but interior to the Oort cloud,
containing trans-Neptunian objects (TNOs) with
$a<10^3$ \AU.  This region contains a complex, overlapping population of
dynamically distinct classes of small bodies.

First there is the classical Edgeworth-Kuiper belt (EKB), a region
estimated to contain a current total mass of the order of 0.01--0.02
$M_\oplus$ (Bernstein et al.\ 2004; Fuentes and Holman 2008).
The observed EKB objects are widely
believed to represent the remains (perhaps less than 1\%) of a
massive primordial population of objects originally formed in low to
moderate-eccentricity orbits in the extended proto-planetary disc beyond
Neptune (Stern 1995, 1996; Morbidelli and Brown 2004).
Non-resonant
EKB objects cannot be the dominant source of observed JF comets as
there are too few observed low-eccentricity orbits in this region with
perihelia close enough to the orbit of Neptune to be captured in sufficient
numbers (see Emel'yanenko et al.\ 2005).  Resonant EKB objects can
diffuse to other dynamical populations over Gyr time-scales (Morbidelli
1997; Tiscareno and Malhotra 2009), but their escape rate is rather
less than that of `scattered disc' objects (Volk and Malhotra 2008),
so that this scattered disc,
a declining and dynamically unstable population introduced by
Duncan and Levison (1997), is a more important source of
JF comets.  For these reasons the classical EKB is not
part of our present model.

A second class of `primordial' TNO (i.e.\ TNOs that
have never reached the Oort
cloud) is a subset of the `scattered disc' population.  In this picture
(Torbett 1989), objects originally formed in the region of the giant
planets are gravitationally scattered outwards in $a$ to produce
an extended, flattened disc-like structure.
Whereas a primordial disc of objects beyond Neptune would be characterized
by low eccentricities and inclinations, according to many theories of
cometary origin, the scattered disc is expected to contain objects on
orbits having much higher eccentricities and substantial inclinations,
perhaps merging smoothly into the unobserved but massive inner Oort cloud
described by Hills (1981).  This second class of TNO therefore comprises
objects that may have encountered Uranus and Neptune during an early phase
of evolution of the Solar system and somehow survived to the present day
without ever having evolved as far as the Oort cloud ($a>10^3$ \AU).  In
our model (Section \ref{model}), for example, 6\% of particles that
had initial perihelion distances in the range $25<q_0<36$ \AU\ survived to
the present day without entering the Oort cloud or reaching any other
end-state of the model (Emel'yanenko et al.\ 2007).
This means that there is likely to be a
significant number of surviving objects in this region
whose orbits would appear to be very long-lived and which
previous work has shown might possibly be a significant source of SP comets
(Duncan and Levison 1997, Emel'yanenko et al.\ 2004).

A third class of TNO comprises bodies that were formed with original
orbits in or close to the proto-planetary disc, but which at some time
in their orbital history became part of the Oort cloud ($a>10^3$ \AU)
and are thus not `primordial' in the sense of the second class above.
Although most objects reaching the Oort
cloud still have $a>10^3$ \AU\ at the present epoch, a few
evolve back to $a<10^3$ \AU\ and so into the trans-Neptunian region.
Our model produces many such objects, which we defined as
`Oort Scattered Disc' (OSD) in Emel'yanenko et al.\ (2007).

We define also the near-Neptune high-eccentricity (NNHE) region, by
$28 < q < 35.5$ \AU\ and $60 < a < 1000$ \AU.
This region has an important dynamical
characterization, covering objects that come close enough to Neptune's
orbit to be captured.  The $q$ cutoff at 28 \AU, just within Neptune's
orbit and below which an object becomes a Centaur, acknowledges the
importance for coming under a planet's control of a particle's
perihelion distance (Horner et al.\ 2003).

{\em Observed\/}
NNHE objects are an important source of SP comets coming from the
trans-Neptunian region (Emel'yanenko et al.\ 2004, 2005). Whether these
observed NNHE objects are the same as NNHE objects produced as a
result of dynamical evolution of objects into and subsequently
from the Oort cloud remains to be
determined.  Section \ref{centaur-constrt} concludes they are not,
and therefore that the observed NNHE objects come from another
source than that considered here.

\section{Integrations}
\label{integ}

\subsection{Model and methods}
\label{model}

To construct our Oort cloud model, following Emel'yanenko
et al.\ (2007), particles' initial conditions after the
formation and migration of the planets had the original
semimajor axes uniformly distributed in the range
$50\!<\!a_0\!<\!300$\,\AU.  The original inclinations were distributed
following a `sine law' scaled to the interval $0\!<\!i_0\!<\!40^\circ$; 
the original perihelion distances were distributed uniformly in the range
$5\!<\!q_0\!<\!36$\,\AU; and the original arguments of perihelion and
original ascending nodes were distributed uniformly between 0 and
$360^\circ$.
The inclination distribution (peaked at $i_0=20^\circ$ falling to zero
at 0 and 40$^\circ$) is similar to the model scattered disc $i$
distribution adopted by Volk and Malhotra (2008) following Brown (2001).
Our choice of $q_0\!<\!36$\,\AU\ is connected with
the assumption that the Oort cloud was created by objects coming from the
planetary region or its nearest vicinity. Although some objects with
$q_0\!>\!36$\,\AU\ may reach the near-Neptune region (Duncan et al.\
1995; Emel'yanenko et al.\ 2003), it is evident that their
contribution to the Oort cloud is small because the rate of diffusion in
perihelion distance is slow.

While our choice of $q_0$ assumes that comets were scattered to the Oort
cloud region mainly by planetary perturbations, we do not use as initial
conditions near-circular orbits in the planetary region (in contrast,
for example, with Dones et al.\ 2004).  Thus although it may be natural
to assume that planetesimals formed in near-circular orbits are a source
of Oort cloud comets, the accretional model of planetary formation still
has so many difficulties and unclear questions that we deliberately
avoid considering any particular hypothesis of comet formation a priori.
Indeed the real situation with the initial orbital distribution of
comets could be much more complicated than that described in Dones et
al.\ (2004) even if comets were formed in near-circular, co-planar
orbits. For example, planetary migration in the early Solar system
appears to have been important in shaping the outer Solar system
(Tsiganis et al.\ 2005). Moreover, the Sun may have formed in a denser
stellar environment than it occupies now (Fern\'andez and Brunini 2000;
Levison et al.\ 2010). This makes assumptions about the distribution of
comets in the early Solar system very uncertain.

Instead our approach is to constrain some features of the cometary
distribution in the early Solar system by analysing observed
distributions of cometary objects in the present Solar system.  The main
aim is to show that there are models of the Oort cloud that
can explain the observed distributions of JF and HT comets.  Our Oort
cloud model can be interpreted as providing some general constraints on
aspects of the cometary orbital distribution during early stages of the
Solar system's evolution.  While details of the
earliest stages of planetary and Oort cloud formation are beyond the
scope of the present paper, we regard our Oort cloud as representing
a general class of model in which cometary planetesimals, formed
in the proto-planetary disc, have been scattered outwards by the planets
to become subject to stellar and Galactic perturbing forces (cf.\
Duncan et al.\ 1987; Fern\'andez 1997; Dones et al.\ 2004;
Dybczy\'nski et al.\ 2008; Leto et al.\ 2008).  Hahn and Malhotra's
(1999) finding (their Section 4) that the total mass reaching the Oort
cloud is quite insensitive to the orbital histories of the migrating
planets tentatively supports our assertion that the precise details of
planetary migration and comet formation are not relevant to our
present purpose.  It is for these reasons that we regard the `initial
conditions' of our integrations as applying to the time after the Solar
system's planetary migration phase.

\begin{figure}
\includegraphics[width=\columnwidth]{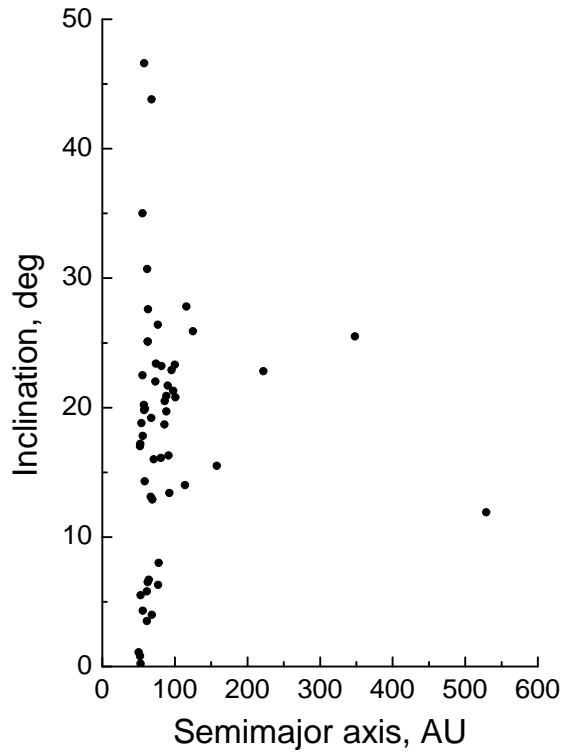}
\caption{The distribution of $a$ and $i$ for all the observed
multiple-opposition high-eccentricity TNOs with $q>36$ \AU.
Data from MPC.}
\label{aiqg36}
\end{figure}

There are several further
motivations for the choice of initial high-eccentricity
($50\!<\!a_0\!<\!300$ \AU; $q_0$ in or near planetary region) rather
than near-circular orbits.  This range of $a_0$ is sufficiently large
that objects can reach it at an early stage of evolution on the way to
the Oort cloud under a wide range of different assumptions of cometary
formation.  The choice of initial conditions also allows particles to
experience planetary perturbations for a long time before reaching the
Oort cloud region, the model's maximum value of $a_0$ being much smaller
than that used in a similar approach by Duncan et al.\ (1987).  The
choice of initial $i$, and initial $a$ ranging above 200~\AU, is
moreover expected in the scattered disc model with migrating Neptune
(Gomes 2003).  The main reason for our choice of initial orbits,
however, is that the majority of high-eccentricity trans-Neptunian
objects have orbits with $50\!<\!a\!<\!300$\,\AU\ and $i<\!40^\circ$.
Figure \ref{aiqg36} shows the distribution of $a$ and $i$ for discovered
multiple-opposition objects with $q>36$ \AU.  This population of
trans-Neptunian objects may preserve at least some memory of its
original early Solar system distribution.  Results for
different initial models can be obtained by applying appropriate
weights (Section \ref{best}).

The initial orbits were integrated in a model Solar system taking full
account of planetary perturbations. All objects that reached the Oort
cloud ($a>10^3$ \AU) were then evolved for the remaining
age of the Solar system under the combined action of planetary, stellar and
Galactic perturbations.

In the present work, the 8925 objects that survived after 4.5 Gyr were
cloned 200 times and integrated for a further 300\,Myr including
planetary, stellar and Galactic perturbations. The initial orbital
distribution of these objects is shown in Figures 1 and 2 of
Emel'yanenko et al.\ (2007). In order to suppress any possible artefacts
associated with the initial conditions of the 300 Myr integrations we
analysed our results on the interval 50--300 Myr. We took account of
perturbations from the four large planets Jupiter to Neptune, using the
secular perturbation theory of Brouwer and van Woerkom (1950) and Sharaf
and Budnikova (1967), adding the terrestrial planets' masses to the Sun.
Objects were removed when $q < 0.005$ \AU\ or $1/a < 10^{-5}$
\AU$^{-1}$, or if they collided with planets.

The orbital calculations used the symplectic integrator described in the
papers Emel'yanenko (2002) and Emel'yanenko et al.\ (2003) unless and
until the orbit reached $q<2.5$ \AU\ and the symplectic integrator of
Emel'yanenko (2007) beyond that. The former method solves the
Hamiltonian equations of barycentric motion for test particles moving in
the field of the Sun and planets. It uses an adaptive time-step that is
a function of the distance $r$ from the centre and of the magnitude of
perturbations, and so can deal with both highly eccentric orbits and
close planetary encounters.  The time-step is almost proportional to $r$
at small distances and in the absence of close encounters: in general it
was 15 days at $r=5$ \AU, and it did not exceed 900 days at any
distance.

For objects reaching $q<2.5$ \AU, the time-step of the integrator was
approximately equal to 4.9989$r$/$\varphi$, where
$\varphi=1+Br+\gamma\sum_{j=1}^4b_j/\Delta_j+\gamma_1/r^{3/2}$,
$B$=0.005549, $\gamma$=3, $\gamma_1$=58, $b_j=a_j(m_j/3)^{1/3}$,
$\Delta_j$ is the distance between the object and the perturbing planet,
and $a_j$ and $m_j$ are the mass and the semimajor axis of the
perturbing planet ($j$=1,2,3,4 for Jupiter, Saturn, Uranus and Neptune,
respectively).

The Galactic model is taken from Byl (1986), but with the Sun's angular
speed $\Omega_0\!=\!26$\,km~s$^{-1}$\,kpc$^{-1}$ and the mid-plane
density of the Galactic disc in the Solar neighbourhood
$\rho_0\!=\!0.1\,M_\odot$~pc$^{-3}$ following Levison et al.\ (2001).
To model stellar perturbations the procedure of Heisler et al.\ (1987)
was used.

\subsection{Initial results}

We have previously shown that objects that have visited the Oort cloud
($a>10^3$ \AU) at some time in their orbital history make a significant
contribution to the observed classes of cometary objects in the Solar
system (Emel'yanenko et al.\ 2007).  Table~2 updates the results of that
work using the present, more extensive simulations, adopting a present-day
near-parabolic flux $\nu_{\mbox{\scriptsize new}}=2.5$.
The difference between the first three
lines of this Table and the corresponding results in Table 2 of
Emel'yanenko et al.\ (2007) are partly due to the assumed
$\nu_{\mbox{\scriptsize new}} = 2$ in
that paper and partly also due to statistical fluctuations in the
relatively small number of objects considered in the earlier work. 

In the present Table 2, $N_{\rm OC}$ is the total number of objects in the
Oort cloud ($a>10^3$ \AU) at the present epoch; and $N_{\rm I}$ and $N_{\rm
O}$ are the corresponding numbers in the relatively flattened inner Oort
cloud ($10^3<a<10^4$ \AU) and the more isotropic outer Oort cloud ($a>10^4$
\AU) respectively.  

$N_{\rm S}$ is the number of OSD objects (objects from the Oort
cloud in the region $q>30$ \AU, $60<a<1000$ \AU, the `S' suffix
indicating that they are located in the analogous region to the
scattered-disc objects discussed by authors such
as Duncan and Levison 1997), $N_{\rm N}$ is the number in the NNHE
region and $N_{\rm C}$ is the number of
Centaurs, also at the present epoch.  In our
model, $N_{\rm S}$, $N_{\rm N}$ and $N_{\rm C}$ represent the
numbers of objects in these respective regions {\em which have
previously visited the Oort cloud\/}.  In order of magnitude, $N_{\rm S}
\approx \mbox{3--4} N_{\rm N}$, the majority in orbits that do not
strongly interact with Neptune, and $N_{\rm N} \approx \mbox{7--8}
N_{\rm C}$.

Finally, $\nu_{\rm JF}$ and $\nu_{\rm HT}$ are the corresponding
present-day annual injection rates of cometary objects coming from the Oort
cloud into JF and HT orbits with $q<1.5$ \AU.   The values
$\nu_{\rm JF}$ and $\nu_{\rm HT}$ are `dynamical' injection rates, i.e.\
obtained by ignoring any effects of physical fading or
disintegration. The total number of active JF and HT comets will depend
(see below) on their respective dynamical and physical lifetimes as SP
comets. It is noteworthy that $\nu_{\rm HT}$ is relatively insensitive to
the initial frequency distribution of objects versus perihelion distance.
Many Halley-types come from long-period Oort cloud orbits with
perihelion distances in the inner planetary region (i.e.\ roughly within
the orbit of Jupiter), but others (roughly 20\% of the total)
originate from the high-eccentricity Oort cloud cometary flux
through the outer planetary region (Emel'yanenko and Bailey 1998;
Emel'yanenko et al.\ 2007) and have a correspondingly more complex
dynamical history.
Some of these comets reaching JF or HT orbits pass through the
$N_{\rm S}$ or $N_{\rm N}$ regions en route from the Oort cloud.

\begin{table}
\begin{center}

\caption{The number of cometary objects in different dynamical classes at
the present epoch.  All figures are calibrated with an assumed
near-parabolic flux $\nu_{\mbox{\scriptsize new}} = 2.5$.
The first three lines show the total
number in the Oort cloud and the contributions to this number from objects
in the inner and outer Oort cloud respectively.  The second three lines
show the numbers of OSD objects ($N_{\rm S}$),  NNHE objects
($N_{\rm N}$) and Centaurs ($N_{\rm C}$)
coming from the Oort cloud.  The final two lines indicate the
present-day rate of  production of new JF and HT comets from this
Oort-cloud source into orbits with $q<1.5$ \AU, neglecting any effects due
to fading.   The columns provide results for four different
frequency distributions of initial perihelion distance,
with relative numbers in the outer Solar system
increasing from left to right.\protect\\
For clarity the dynamical definitions used for these classes are then
summarized. The $N_{\rm S}$ and $N_{\rm N}$ classes overlap; we
primarily use $N_{\rm N}$ to analyse data (see especially Section
\ref{centaur-constrt}) but calculate $N_{\rm S}$ for extra comparisons
with other work.}
\label{noc}

\begin{tabular}
{@{}c r@{$\times$}l r@{$\times$}l r@{$\times$}l r@{$\times$}l @{}}
\hline
 \\
            & \mc{$q^{-2}$} & \mc{$q^{-1}$} & \mc{$q^0$}    & \mc{$25<q<36$}
\\[5pt]
$N_{\rm OC}$& 4.8&$10^{11}$ & 5.3&$10^{11}$ & 5.8&$10^{11}$ & 7.1&$10^{11}$\\
$N_{\rm I}$ & 1.7&$10^{11}$ & 2.2&$10^{11}$ & 2.6&$10^{11}$ & 4.1&$10^{11}$\\
$N_{\rm O}$ & 3.1&$10^{11}$ & 3.1&$10^{11}$ & 3.1&$10^{11}$ & 3.0&$10^{11}$
\\[5pt]
$N_{\rm S}$ & 9.0&$10^{9}$  &18.0&$10^{9}$  &21.0&$10^{9}$  &43.0&$10^{9}$\\
$N_{\rm N}$ & 3.0&$10^{9}$  & 5.6&$10^{9}$  & 6.5&$10^{9}$  &12.9&$10^{9}$\\
$N_{\rm C}$ & 4.6&$10^{8}$  & 7.7&$10^{8}$  & 8.4&$10^{8}$  &15.2&$10^{8}$
\\[5pt]
$\nu_{\rm JF}$ & \mc{0.043} & \mc{0.069} & \mc{0.100} & \mc{0.203} \\ 
$\nu_{\rm HT}$ & \mc{0.073} & \mc{0.079} & \mc{0.082} & \mc{0.083} \\[5pt]
\hline\\
$N_{\rm OC}$   & \multicolumn{8}{l}{$a>1000$} \\
$N_{\rm I}$    & \multicolumn{8}{l}{$1000<a<10000$} \\
$N_{\rm O}$    & \multicolumn{8}{l}{$a>10000$}
\\[5pt]
$N_{\rm S}$    & \multicolumn{8}{l}{$q>30,\;\;60<a<1000$} \\
$N_{\rm N}$    & \multicolumn{8}{l}{$28<q<35.5,\;\;60<a<1000$} \\
$N_{\rm C}$    & \multicolumn{8}{l}{$5\!<\!q\!<\!28,\;\;a\!<\!1000$
                                    (not resonant TNOs, Trojans)}
\\[5pt]
$\nu_{\rm JF}$ & \multicolumn{8}{l}{$P<200,\;\;T>2$} \\
$\nu_{\rm HT}$ & \multicolumn{8}{l}{$P<200,\;\;T<2$} \\[5pt]
\hline
\end{tabular}

\end{center}
\end{table}

\section{Short-period comet problems}

\subsection{Numbers}
\label{num-prob}

It is well known that, with a population of only $\sim$100 HT comets with
$q<1.5$ \AU\ (as constrained by observations), if we try to explain their
origin by capture from the present-day Oort-cloud near-parabolic flux with
initial perihelion distances $q_{\mbox{\scriptsize init}}<5$ \AU,
then it is necessary to place a
very tight limit on the physical lifetime of such comets.  This limit is
further strengthened by the inclusion of HT comets originating from
Oort-cloud source orbits with initial perihelion distances
$q_{\mbox{\scriptsize init}}>5$ \AU.  Since comets
are typically active at larger distances than 1.5 \AU, we must also
consider restrictions on their physical lifetime in the region
$q<2.5$~\AU.  Thus, for particles reaching $q<1.5$~\AU, our integrations
record also the preceding length of time spent with $q<2.5$ \AU.

Although highly volatile ices, such as carbon monoxide CO, can sublimate at
large distances $\sim$10 \AU, the main driver of cometary activity, as
recognized long ago by Whipple (1950), is the sublimation of water H$_2$O
ice. The mass loss rate for sublimating water ice has a fast decrease for
heliocentric distances larger than 2 \AU\ (Jewitt 2004). Therefore, in our
model we apply restrictions on the cometary lifetime only in the region
$q<2.5$ \AU, assuming that outside this region the fading of comets is
negligible in comparison to that when $q<2.5$ \AU.

In order that the steady-state number of active HT comets should be
$\ltsimeq$100, our
results imply that objects from the Oort cloud ($a>10^3$ \AU\ at some time
during their history) should survive as active comets for an average of
$\ltsimeq$150 revolutions in the region $q<2.5$ \AU, in
the model where the number of objects per unit perihelion distance
is proportional to $q_0^{-2}$.  The result is much the same for other
models, as indicated by the relatively weak dependence of $\nu_{\rm HT}$
versus dynamical model given in Table~2.

However, when we apply the same physical-lifetime constraint to the
Oort-cloud objects that eventually become JF comets, we
predict too few JF comets by a factor of around 30. That is, we predict
only about three JF comets in the region $q<1.5$ \AU\ compared to the
$\sim$100 to be explained.  This illustrates the well-known problem of
explaining the
number of JF comets captured from the Oort cloud if the two classes of SP
comet are assumed to have broadly the same physical properties and
lifetimes, a result (as we have indicated) at the heart of
what we have called the SP comet fading problem.

There is an extensive literature on possible ways to overcome this `number'
problem, including the assumption that JF
comets may arise through the fragmentation of one or more large
progenitors.   Such time-dependence in the present-day JF
population appears to be rather unlikely, and in recent years has
led to an increasing focus on models in which not only do JF and HT
short-period comets originate from different primordial source regions,
but have different physical properties as well.

\subsection{Fading}

The problem of the relative and absolute numbers of HT and JF comets
suggests the need to introduce an alternative dominant source for JF comets
other than the Oort cloud. Such a source could include a remnant population
of scattered-disc objects perturbed by Neptune on to relatively long-period
orbits at an early stage of Solar-system evolution (Duncan and Levison
1997), or for example a primordial population of high-eccentricity
trans-Neptunian objects initially formed beyond Neptune
(or a combination of these pictures). However, irrespective of
the details of such a model, there would remain the SP comet fading
problem.  That is, the problem that JF comets originating from any
such trans-Neptunian source
must have much longer lifetimes in the inner Solar system than observed
Halley-types, and therefore statistically different fading properties.

The problem can be illustrated in three ways.  First, Emel'yanenko et al.\
(2004) showed that, while the outflow rate from the observed
NNHE region is $\sim$$0.93 \times 10^{-9}$ yr$^{-1}$
(a factor of 3 higher than the outflow rate found by Volk and
Malhotra (2008) from the more limited region having $q>33$ \AU),
the predicted dynamical injection rate of JF comets with
$q<1.5$ \AU\ from the observed NNHE region is approximately $0.18 \times
10^{-10} N'_{\rm N}$ yr$^{-1}$. Here $N'_{\rm N}$ is the intrinsic (i.e.\
observationally debiased) number of objects in the NNHE region represented
by the then observed sample.

If it is assumed that most JF
comets come from this region and have broadly the same fading
behaviour as the observed Halley-types (i.e.\ mean lifetimes of the order
of 150 revolutions in the region $q<2.5$ \AU), then our calculations
would require $N'_{\rm N} > 3 \times 10^{10}$.  
This value, which as in Section \ref{obs-com} may be assumed to apply to
$H_{10} \ltsimeq 11$ cometary bodies (nuclear diameter $\gtsimeq$ 1 km),
is greater than all previous
estimates of the number of objects in this NNHE region, for example the
$\sim\!4 \times 10^9$ scattered-disc objects with $q$ in the range
34--36~\AU\ estimated by Trujillo et al.\ (2000).  Furthermore, our
result is a lower limit, for example because some of the comets that might
have reached $q<1.5$~\AU\ in the absence of fading will be removed from the
distribution of active comets by the lifetime limit of 150 revolutions
within $q < 2.5$~\AU.  Thus, if we take account of physical fading, the
rate of injection of active JF comets to the region $q<1.5$~\AU\ is less
than $0.18 \times 10^{-10} N'_{\rm N}$ yr$^{-1}$, requiring an even larger
number of objects in the observed NNHE region to explain the observed
number of JF comets.

A second argument comes from the predicted inclination distribution of the
resulting JF comets.  Emel'yanenko et al.\ (2004) found that the observed
JF comets could in principle be explained by the evolution of objects
captured from the observed NNHE region provided that the maximum lifetime
of the resulting JF comets in the region $q<2.5$~\AU\ was not too long,
i.e.\ approximately 2500 years ($\sim$360 revolutions).  However, such a
lifetime (i.e.\ 360 revolutions) is already 2--3 times longer than that
required to explain the active HT comets from our Oort-cloud source, again
highlighting the SP comet fading problem.

A third general argument leading to the same conclusion arises because the
estimates in Emel'yanenko et al.\ (2004) were based on the assumptions that
(a) the number of objects in the observed NNHE region is of the order of
$10^{10}$ and (b) the physical behaviour of all JF comets is broadly the
same.  If the number of objects in the observed NNHE region is smaller than
this, as seems likely (e.g.\ Levison et al.\ 2006), we would have to invoke
longer average JF comet lifetimes in the region $q<2.5$ \AU\ to explain the
observed number.  Alternatively, if there are two types of JF comet, for
example one with a mean lifetime in the region $q<2.5$ \AU\ comparable to
that ($\sim$150 revolutions) required to explain the number of HT comets,
then the other must have a much longer average lifetime to compensate. 
This would exacerbate the SP comet fading problem, not just by highlighting
a difference in the physical properties of some JF comets and Halley-types,
but by introducing a new (and arbitrary) difference between two
different assumed types of JF comet.

Various arguments could of course be invoked to justify possible physical
differences between different types of comet, for example that Oort-cloud
objects might have visited the Jupiter-Saturn region many times before
being finally ejected into the outer Solar system, whereas NNHE objects
might never have come close to the inner Solar system before finally
evolving into the observed Jupiter family. In this case,
and whatever one's view of the merit of such speculations,
it is evident that we should not dismiss lightly
the possibility that there may be two or more distinct types of SP
comet.

However, in order to accommodate the twin constraints of the number of JF
comets (tending to require a long lifetime) and their inclination
distribution (tending to require a shorter lifetime), such models are also
subject to fine tuning and a very strict observational test.  That is, the
dynamically distinct HT and JF classes of SP comet should, on average, have
very different fading properties and rates of decay in the observable
region.  In particular, the JF comets originating from a flattened source
distribution other than the Oort cloud must, if they are to dominate the
observed distribution of JF comets, have much longer lifetimes in the
observable region than their HT counterparts originating from the Oort
cloud.  In principle, such a major physical difference between the two
dynamical classes of SP comet (or even within the dynamically defined
Jupiter family if the latter come from originally distinct sources) should
be amenable to an observational test.

In summary, the SP comet fading problem remains an obstacle to
understanding the origin of SP comets. Although it may be reasonable to
suggest that the comets which formed in different regions of the primordial
Solar system might have different fading properties after they eventually
evolve into the observable region, it is important
to emphasize that there is as yet no clear-cut
observational evidence to support such a view, nor even for any clear
physical differences between the two main classes of SP comet.  Rather than
two physically different types of SP comet, behaving in statistically
different ways in the inner planetary region so far as fading is concerned,
we therefore instead develop in the remainder
of this paper a unified model for the
origin of SP comets.  In this unified model all comets, whether coming from
the Oort cloud or trans-Neptunian region, display broadly similar physical
behaviour in the inner planetary region.

\section{Unified model}

We return to the idea that the key factor linking the two classes of SP
comet, and perhaps all classes of comet, is their singular lack of strength
and associated rapid fading.  We thus seek a unified model for the origin
and evolution of cometary bodies in the Solar system (particularly the
observed SP comets) in which the majority of observed SP comets (though
perhaps not all) originate from an Oort-cloud source which itself has an
origin primarily in the dynamical evolution of objects left behind after
the period of planet formation and planetary migration.  In this case it is
reasonable to assume that, to first order, the majority of comets will have
broadly similar characteristics, though not necessarily identical physical
properties, including those relating to fading.

In developing this unified physical picture for the origin of comets, we
obtain new constraints on their required fading properties within the
observable region.  In particular, we use dynamical information provided by
the results of our integrations and the link between Centaurs and SP comets
to constrain the cometary numbers and lifetimes. In broad terms, our
unified model predicts that essentially all the HT comets and nearly half
the JF comets come from the Oort cloud. A flattened trans-Neptunian
disc source is, however, required for the remaining $\sim$50\% of JF
comets.  However, these objects too are predicted to have relatively short
physical lifetimes within the observable region in order not to produce too
many active JF comets. Thus, all comets have essentially the same fading
properties within the observable region.

\subsection{Centaurs and the NNHE region}
\label{centaur-constrt}

In principle, understanding the relative contributions of different
outer Solar system source regions to the SP comet population requires
a full description of the number and orbital distribution
of all objects in the trans-Neptunian region.  Unfortunately our
present knowledge of this complex region is limited by the precision
with which the observed orbits are known and by severe observational
selection effects. We therefore use the observed distribution of Centaurs
(objects with $5<q<28$ \AU\ and $a< 1000$ \AU) to constrain our results. 
Centaurs are an important transition population providing valuable
information. Emel'yanenko (2005) and Emel'yanenko et al.\ (2007)
presented various characteristics of the orbital distribution of
Centaurs from the Oort cloud, results which are supported by our
present work. The more extensive integrations of our
current paper are necessary to provide a sufficient number of integrated
particles transferred from the outer Solar system to SP orbits.

We recall that Emel'yanenko et al.\ (2005) predicted the orbital
distribution of Centaurs originating from the observed NNHE region
($28<q<35.5$ \AU\ and $60<a<1000$ \AU).  These early results were based on
the orbits of seven well-determined observed TNOs in the NNHE region
suitably weighted by an observational debiasing procedure (Emel'yanenko et
al.\ 2004).  Let us denote as $N'_{\rm N}$ the intrinsic (i.e.\ debiased)
number of objects in the NNHE region represented by this observed sample. 
Note that $N_{\rm N}$ introduced above in Section \ref{tno} is defined in
terms of exactly the same region of orbital element phase space.  However,
whereas $N_{\rm N}$ refers to objects that have been in the Oort cloud ($a>
10^3$~\AU), $N'_{\rm N}$ is the intrinsic (observationally debiased) number
of NNHE objects represented by the discovered population.  By this
definition, $N'_{\rm N}$ and $N_{\rm N}$ could comprise the same
population, or be disjoint, or partially overlap.  If disjoint, then
$N'_{\rm N}$ could represent the number of objects in the NNHE region
associated with a primordial source distribution in the trans-Neptunian
disc and so not be included in our Oort-cloud model.  We can discover how
$N'_{\rm N}$ really relates to $N_{\rm N}$ by using Centaurs as a
constraint, as follows.

In a steady state, the number of Centaurs $N'_{\rm C}$ originating from the
observed NNHE source region is a fixed proportion of the total number
$N'_{\rm N}$ of such objects.  Emel'yanenko et al.\ (2005), using the
integrations of Emel'yanenko et al.\ (2004), calculated the constant of
proportionality $f_{{\rm N'}\rightarrow{\rm C'}} \simeq 0.008$, i.e.\
$N'_{\rm C} \simeq 0.008N'_{\rm N}$.  They also showed that these Centaurs
were split in the ratio 0.003 to 0.005 between orbits having respectively
$a>60$ and $a<60$ \AU, nearly all the latter having $20<a<60$ \AU\
(Emel'yanenko et al.\ 2005, fig.~2).

In order to compare these dynamical results with observations it is
necessary to apply an appropriate debiasing correction to the observed
distribution of Centaurs.  The results of Emel'yanenko et al.\ (2005),
based on a sample of 42 well-determined Centaur orbits excluding objects in
the 2/3 mean-motion resonance with Neptune, showed that the intrinsic
number of Centaurs $N^{\rm obs}_{\rm C}$ is overwhelmingly dominated by
objects with $a>60$ \AU\ (roughly 90\% of Centaurs having such
orbits), and that $N_{\rm C}^{\rm obs} \simeq 0.13N'_{\rm N}$.  This ratio,
namely 0.13, is much larger than the dynamical prediction
$f_{{\rm N'}\rightarrow{\rm C'}} \simeq 0.008$, and this fact alone
implies that the majority of Centaurs, particularly the majority of
those with $a>60$ \AU, must have another source, i.e.\ a source other
than the $N'_{\rm N}$ objects representing the observed NNHE region.
In this case, because it is an inescapable part of any successful model,
such a source is most likely the Oort cloud.

Emel'yanenko et al.\ (2005) also showed (their fig.~5) that, after
debiasing, only 10\% of Centaurs with $a<60$~\AU\ have $40 < a <
60$~\AU.  On the other hand, if the principal source of Centaurs had been
the observed NNHE region, the dynamically predicted fraction would have
been around 50\% (loc.\ cit.\ fig.~2).  This is further evidence
that the $N'_{\rm N}$ objects representing the observed NNHE region cannot
explain all the observed Centaurs.  Indeed, it raises the possibility that
the Oort cloud may contribute significantly to Centaurs with $a<60$ \AU\ as
well.

In summary, the dynamically predicted number of Centaurs with $a>60$
\AU\ coming from the observed NNHE region is roughly $0.003 N'_{\rm N}$,
whereas observations require this number to be of the order of 90\%\,$\times
0.13 = 0.117N'_{\rm N}$.  The difference between these two results (i.e.\
$0.114 N'_{\rm N}$) can be attributed to an Oort-cloud flux, i.e.\ the flux
of Oort-cloud objects through the planetary system irrespective of whether
they have gone through the NNHE region. At this stage we make no assumption
as to whether any or all of the $N'_{\rm N}$ objects represented by the
observed NNHE population come from the Oort cloud.  In any case, their
contribution to Centaurs with $a>60$ \AU, i.e.\ $\simeq 0.003 N'_{\rm N}$, 
is insignificant.

Our new integrations provide a value for the steady-state ratio of the
number of Centaurs produced from the Oort cloud with $a<60$~\AU\ to the
number with $a> 60$~\AU\ (cf.\ Table~\ref{nmod} later).
Specifically, for every
Centaur with $a>60$~\AU, approximately 0.07 Centaurs are produced with
$a<60$~\AU. Therefore, for every $0.114N'_{\rm N}$ Centaurs with $a>60$
\AU\ that the Oort cloud produces, it also produces $\sim \! 0.008 N'_{\rm
N}$ Centaurs with $a<60$ \AU. 

As we have noted, the dynamically predicted number of Centaurs with $a<60$
\AU\ coming from the observed NNHE region is $N'_{\rm C}(a<60) \simeq 0.005
N'_{\rm N}$ and the debiased number of Centaurs with $a<60$ \AU\ is $N_{\rm
C}^{\rm obs} (a<60) \simeq 10$\%\,$\times 0.13 N'_{\rm N} \simeq
0.013N'_{\rm N}$.  Thus, the additional population of Centaurs with
$a<60$~\AU\ produced by the Oort-cloud flux through the planetary system is
sufficient to account for this difference of $0.008N'_{\rm N}$.  However,
to a good approximation, the same Oort cloud flux does not explain the
entire number of $N_{\rm C}^{\rm obs} (a<60)\simeq 0.013N'_{\rm N}$
Centaurs with $a<60$~\AU, the $0.005N'_{\rm N}$ objects from the observed
NNHE region being unaccounted for.   

We conclude that the observed $N'_{\rm N}$ objects are not produced from
the Oort cloud.  In other words, the observed NNHE objects studied by
Emel'yanenko et al.\ (2004) illustrate the dynamical features of
near-Neptune high-eccentricity objects that have never visited the Oort
cloud.  In contrast, the predicted $N_{\rm N}$ NNHE objects originating from
the Oort cloud in our model represent a sample of objects which owing to
discovery biases are under-represented in the observed population.

Thus, although we defined $N'_{\rm N}$ in terms of the observationally
debiased known population, we may now interpret it as referring to a
`primordial' trans-Neptunian population that has never become part of the
Oort cloud ($a > 10^3$ \AU). So while the numbers $N_{\rm N}$ and
$N'_{\rm N}$ describe objects in the same region of orbital element
phase space, they are essentially disjoint sets of objects. The
$N_{\rm N}$ objects coming from a proximate source in the Oort cloud
are largely unobserved, i.e.\ are not yet represented in the
$N'_{\rm N}$ population of observed objects in the NNHE region.

These results allow us to estimate the number $N'_{\rm N}$ of NNHE objects
that have never visited the Oort cloud.  Thus, because the two sources are
disjoint, $N_{\rm C}^{\rm obs} = N_{\rm C} + N'_{\rm C}$, and hence $N_{\rm
C} = 0.122 N'_{\rm N}$ where $N_{\rm C}$ is listed in Table \ref{noc}. 
This in turn allows us to determine the additional contribution of these
`primordial' NNHE objects to the number of Centaurs ($N'_{\rm C} = 0.008
N'_{\rm N}$) and to the flux $\nu_{\rm JF}$ of JF comets with $q<1.5$~\AU,
taking $\nu'_{\rm JF}/N'_{\rm N}=0.18 \times 10^{-10}$ from Emel'yanenko et
al.\ (2004).   

These values are given in Table \ref{primordial} for the same distributions
of initial $q_0$ as in Table \ref{noc}. As with Table \ref{noc}, $\nu'_{\rm
JF}$ is a `dynamical' annual injection rate, i.e.\ assuming no physical
lifetime limit.  For comparison, the scattered disc proposed as a source of
JF comets by Duncan and Levison (1997) corresponds to objects whose
evolution was dominated by initial close encounters with Neptune during the
early dynamical history of the Solar system, with no restriction on their
subsequent evolution in semimajor axis. What we term the `primordial' NNHE
region overlaps this scattered disc to a large extent but does not include
objects that ever reached $a > 10^3$ \AU.

\begin{table}
\begin{center}

\caption{The number of Centaurs $N'_{\rm C}$ and the annual injection rate
of JF comets $\nu'_{\rm JF}$ coming from the observed $N'_{\rm N}$
population of primordial `trans-Neptunian' (TN) NNHE objects.  The columns
correspond to four models as in Table \ref{noc}.  Note that the intrinsic
number of primordial objects in this region, $N'_{\rm N}$, is comparable in
order of magnitude to the number, $N_{\rm N}$, in the NNHE region that
come from the Oort cloud (cf.\ Table \ref{noc}) and which are still
largely undiscovered.  Similarly the annual dynamical injection rate
$\nu'_{\rm JF}$ of JF comets from this primordial TN region is comparable
to the rate $\nu_{\rm JF}$ from the Oort cloud (Table \ref{noc}).  On the
other hand, the predicted number of Centaurs coming from this observed NNHE
TN region is roughly an order of magnitude smaller than that coming
from the Oort cloud.} \label{primordial}

\begin{tabular}{ccccc}
\hline
 & & & & \\
 & $q_0^{-2}$ & $q_0^{-1}$ & $q_0^0$ & $25<q_0<36$\,\AU\ \\[5pt]

$N'_{\rm N}$ & $3.7 \times 10^{9}$ & $6.3 \times 10^{9}$  & $6.9 \times
10^{9}$  & $12.5 \times 10^{9}$ \\

$N'_{\rm C}$ & $3.0 \times 10^{7}$& $5.0 \times 10^{7}$ & $6.0 \times
10^{7}$   & $10.0 \times 10^{7}$ \\

$\nu'_{\rm JF}$ & 0.067    & 0.113  & 0.124 & 0.225 \\[5pt]
\hline
\end{tabular}

\end{center}
\end{table}

\subsection{Initial perihelion distribution}
\label{init-orb}

A further important factor that allows us to discover features of the
dynamical and physical evolution of comets is the orbital distribution of
JF comets. In particular, the predicted distribution of inclinations is very
sensitive to the physical lifetime of comets (Levison and Duncan 1997). On
this basis, we obtained limits of 2500 yr for the physical lifetime of JF
comets in the region $q<2.5$ \AU\ and 1200 yr in the region $q<1.5$ \AU\
(Emel'yanenko et al.\ 2004), assuming all JF comets come from the
trans-Neptunian region. In our present calculations, we have found that JF
comets coming from the Oort cloud have similar dynamical characteristics
and that the modelled $i$ distribution of JF comets is close to the
observed $i$ distribution if the above physical lifetime limits are
imposed.

But if we impose these limits on all SP comets, we have the problem of
numbers described above (Section \ref{num-prob}):  the resulting ratio
of the number of HT to JF comets is too large.  From observational
constraints, this ratio is around 1 -- maybe below 1 but unlikely to
be more than 1.5 (Section \ref{obs-sp}).
We find the ratio ranges from 3.2 for the Oort-cloud model with
initial perihelia within $25<q_0<36$ \AU\ to 12.3 for the model with the
initial distribution proportional to $q_0^{-2}$. In addition, the absolute
number of JF comets is too small in models where objects are initially
concentrated towards lower $q_0$, e.g.\ the number is only 12 in the case
of the $q_0^{-2}$ distribution. An additional SP comet contribution from
the `primordial' trans-Neptunian region does not solve these difficulties:
adding these SP comets (based on the data of Table \ref{primordial} but
with the physical lifetime limits imposed) the HT/JF ratio ranges
from 1.5 to 4.6, the number of JF comets being 32 for the $q_0^{-2}$
distribution.  Overall these constraints favour models where the initial
number of objects increases with $q_0$ and are against models where the
number decreases with $q_0$.

\subsection{Best-fitting models}
\label{best}

In order, therefore, to explore a suitable family of models, we assume
firstly that the initial number of objects versus perihelion distance
follows a power-law distribution, i.e.\ the number of objects in the
range $(q_0, q_0+dq_0)$ is proportional to $q_0^\alpha dq_0$.  To obtain
consistency with both the numbers and orbital distributions of observed
SP comets we also introduce a model for the physical lifetime in the
observable region $q<2.5$ \AU.  Protoplanetary disc models suggest the
snow line (boundary beyond which ice can condense) gradually moves
inwards from distant regions (Davis 2005; Ciesla and Cuzzi 2006; Garaud
and Lin 2007; Oka et al.\ 2011; Martin and Livio 2012) implying that the
water distribution in the early Solar system would have been a function
of heliocentric distance. It follows that comets' composition could
depend on their initial perihelion distance $q_0$ in the early Solar
system. We assume the physical lifetime -- within $q<2.5$ \AU\ for
comets that reach this region at the present epoch -- is a constant
number $n_2$ of revolutions for all objects formed in the outer $q_0$
range (25,36) \AU\ and varies as $q_0^{\beta}$ for $q_0<25$ \AU\ (with
no discontinuity at $q_0=25$). We impose an equivalent restriction, with
the same $\beta$, for the lifetime in the region $q<1.5$ \AU\ at the
present epoch, i.e.\ $n_1$ revolutions when the initial $q_0$ is within
(25,36) \AU\ and $n_1 (q_0/25)^\beta$ for $q_0<25$ \AU.

We have explored which values of these four parameters $\alpha$,
$\beta$, $n_1$ and $n_2$ are consistent with the observational
constraints.  The total steady-state number of JF comets
(to be compared to the number derived from observations) is a sum of the
$N_{JF}$ which we calculate here, originating from the Oort cloud, and the
additional contribution $N'_{JF}$ from the `primordial' trans-Neptunian
population.  $N'_{JF}$ ranges from $\sim$50 for $\alpha=1$ to $\sim$70 for the
model where objects are initially concentrated in the outer region
$25<q_0<36$ \AU.

As we saw (Section \ref{init-orb}), models with $\alpha < 0$ produce
unsatisfactory results, namely too few JF comets as well as an incorrect
value for the HT/JF ratio.  Thus $\alpha>0$ is implied, i.e.\ a greater
initial concentration of comets towards larger initial $q_0$.  Moreover
for values of $\alpha$ larger than 2 (i.e.\ a strong initial concentration of
comets towards the outer region), we need to introduce very strict
restrictions on the cometary lifetime, and the resulting number of HT
comets in retrograde orbits becomes too small in comparison with the
observed number.

Our calculations show that models with $\beta \ge 1$ give results close
to observations.  But provided $\beta \gtsimeq 1$, it is less tightly
constrained than $\alpha$ and can even increase to infinity (formally
$\beta = \infty$ means that all comets that do not originate within the
outer region $25<q_0<36$ \AU\ die after the first perihelion passage
with $q<2.5$ \AU).

Overall it is impossible to derive unique constraints on the cometary
lifetimes and the values of $\alpha$ and $\beta$ simultaneously because
of uncertainties in the number and the orbital distribution of SP comets.
A range of possible solutions for $N_{JF}$ and $N_{HT}$ is presented in
Table \ref{dmod}, representative of the allowed combinations of
parameters $\alpha$, $\beta$, $n_1$ and $n_2$.  The best solutions
correspond to a lifetime limit $n_1 \approx 150$ revolutions, and $n_2
\approx 400$ revolutions, with $\alpha$ being in the approximate range 1
to 2, although there are other possibilities (e.g.\ the first solution
in Table \ref{dmod}) with $n_1$ or $n_2$ differing by up to a few tens of
per cent.

\begin{table}
\begin{center}

\caption{The number of JF and HT comets for various acceptable
combinations of the parameters $\alpha$, $\beta$, $n_1$ and $n_2$.}
\label{dmod}

\begin{tabular}{cccccc}
\hline
 & & & & & \\
$\alpha$ & $\beta$ & $n_1$ & $n_2$ & $N_{JF}$ & $N_{HT}$ \\[5pt]
0.5 & $\infty$ & 170 & 600 & 45 & 112 \\
1   & 1        & 150 & 420 & 42 & 118 \\
1   & 2        & 150 & 420 & 42 & 108 \\
1   & $\infty$ & 150 & 420 & 41 & 101 \\
2   & 2        & 140 & 400 & 46 & 112 \\
2   & $\infty$ & 140 & 400 & 45 & 107 \\[5pt]
\hline
\end{tabular}

\end{center}
\end{table}

\begin{table}
\begin{center}

\caption{One of the best-fitting models.
The number of cometary objects evolving to
different dynamical classes from various initial ranges of $q_0$ (for Oort
cloud comets) and from the observed $N'_{\rm N}$ population of `primordial'
trans-Neptunian (TN) NNHE objects. Here $\alpha =1$ and $\beta=2$; the
restrictions $n_2 = 420$ and $n_1 = 150$ revolutions are used when
calculating $N_{JF}$, $N'_{JF}$ and $N_{HT}$.  $\bar{\nu}$ denotes the
contribution to the observed near-parabolic flux, $\nu_{\mbox{\scriptsize
new}}$, from comets originating respectively in each of the initial ranges
of perihelion distance.} \label{nmod}

\begin{tabular}
{@{}c r@{$\times$}l r@{$\times$}l r@{$\times$}l r@{$\times$}l @{}}
\hline
\\[-5pt]
Initial region: & \mc{5--10 \AU} & \mc{10--25 \AU} &
\mc{25--36 \AU} & \mc{TN} \\[5pt]

$N_{\rm OC}$ & 1.0&$10^{9}$  & 1.8&$10^{11}$ & 4.3&$10^{11}$ & \mc{--}\\
$N_{\rm I}$  & 8.0&$10^{7}$  & 5.0&$10^{10}$ & 2.5&$10^{11}$ & \mc{--}\\
$N_{\rm O}$  & 1.0&$10^{9}$  & 1.3&$10^{11}$ & 1.8&$10^{11}$ & \mc{--}\\[5pt]

$N_{\rm S}$  & \mc{0}        & 1.0&$10^{9}$ & 2.6&$10^{10}$ & \mc{--}\\
$N_{\rm N}$  & \mc{0}        & 3.0&$10^{8}$  & 7.9&$10^{9}$ & \mc{--} \\
$N_{\rm C}$  & 3.0&$10^{4}$  & 9.0&$10^{7}$  & 9.3&$10^{8}$ & \mc{--} \\
$N_{\rm C} (a<60)$ & \mc{0}  & 1.0&$10^{6}$  & 6.6&$10^{7}$ & \mc{--} \\[5pt]

$N'_{\rm N}$ & \mc{--} & \mc{--}  & \mc{--} & 8.3&$10^{9}$\\ 
$N'_{\rm C}$ & \mc{--} & \mc{--}  & \mc{--} & 6.6&$10^{7}$\\ 
$N'_{\rm C}(a<60)$ &\mc{--} & \mc{--} & \mc{--} & 4.2&$10^{7}$\\[5pt]

$\bar{\nu}$  & \mc{0.01}& \mc{0.97} & \mc{1.52} & \mc{--} \\
$N_{JF}$     & \mc{0}   & \mc{1}    & \mc{41}   & \mc{--} \\
$N'_{JF}$    & \mc{--}  & \mc{--}   & \mc{--}   & \mc{51} \\
$N_{HT}$     & \mc{0}   & \mc{7}    & \mc{101}  & \mc{0}  \\[5pt]

\hline
\end{tabular}

\end{center}
\end{table}

\begin{figure}
\includegraphics[width=\columnwidth]{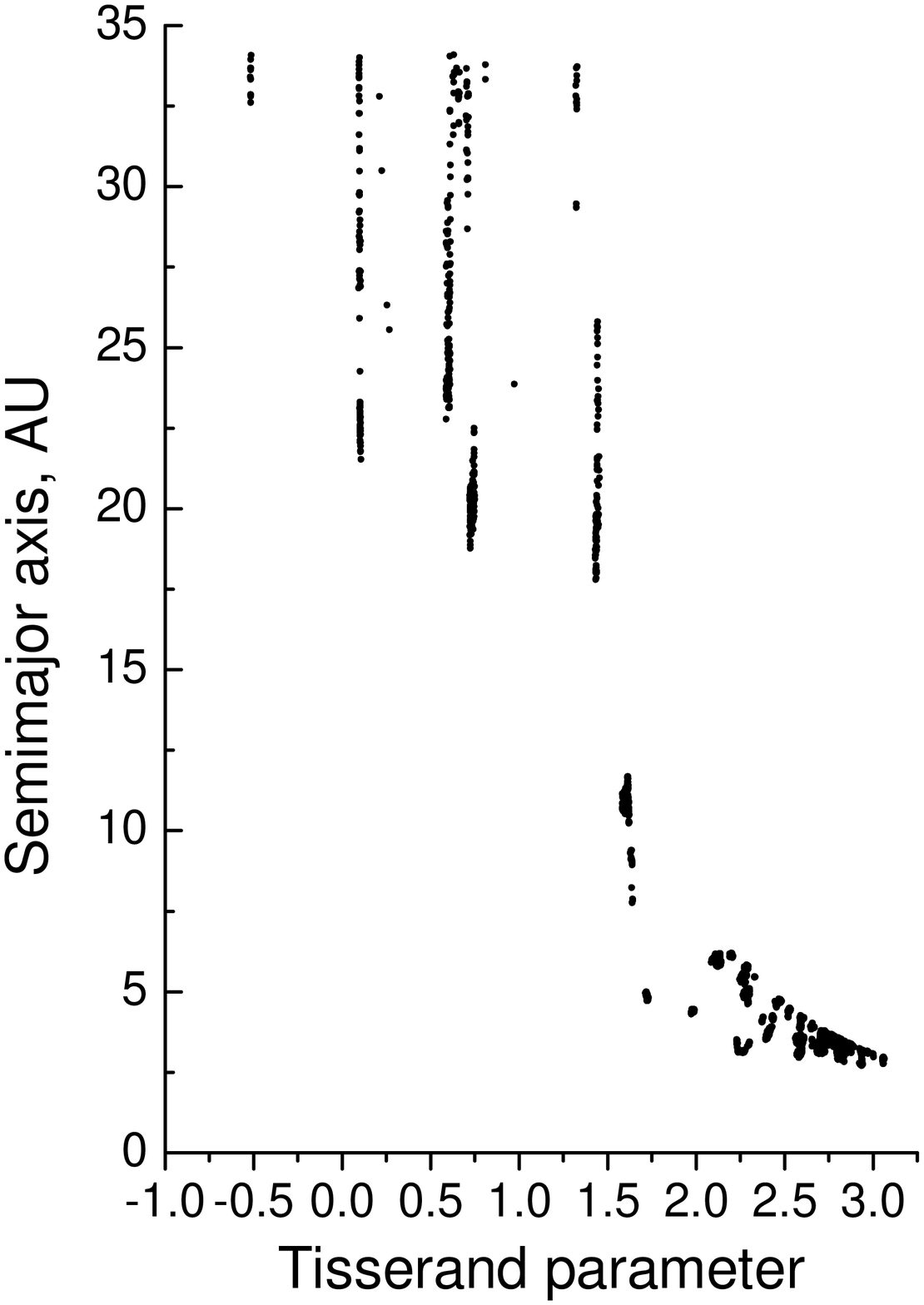}
\caption{The model distribution of $T$ and $a$ in perihelia
for SP comets with $q<1.5$ \AU\ coming from the Oort cloud.}
\label{atall}
\end{figure}

\begin{figure}
\includegraphics[width=\columnwidth]{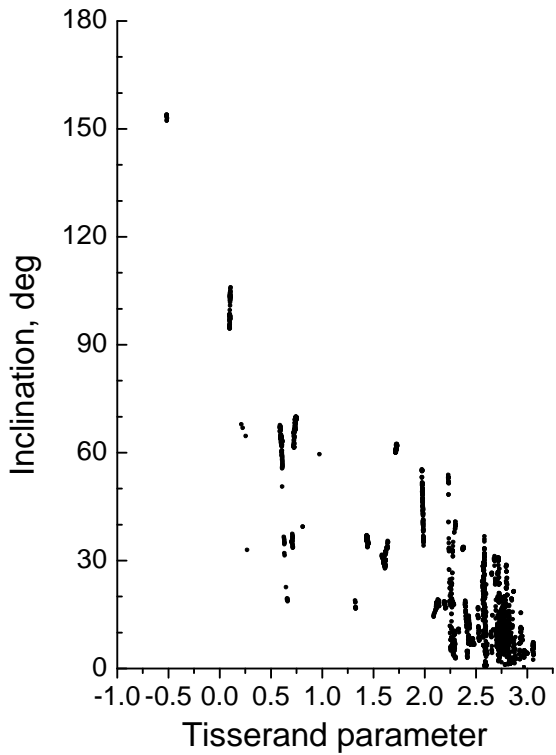}
\caption{The model distribution of $T$ and $i$ in perihelia
for SP comets with $q<1.5$ \AU\ coming from the Oort cloud.}
\label{itall}
\end{figure}

Table~\ref{nmod} summarizes our results for one of the best-fitting
models.  The parameters are $\alpha=1$, $\beta=2$, $n_1 = 150$ and
$n_2 = 420$.  Our model is consistent with the observed features of SP
comets, Centaurs and TNOs, and Table~\ref{nmod}
estimates the numbers of present-day cometary objects coming from the
various original source regions.
The regions correspond to three ranges of initial $q_0$
for objects that have visited the Oort cloud, with objects originating from
the `primordial' trans-Neptunian population that have never been in the Oort
cloud listed in the final column (TN).  The notation in Table~\ref{nmod} is
as introduced earlier, with data also listed for the subset of Centaurs
having $a<60$ \AU.

Whereas the model strongly constrains the initial $q_0$ distribution,
results are not highly sensitive to the initial $a_0$ distribution,
adopted as uniform in the range 50--300 \AU. For example, for the
best-fitting model of Table~\ref{nmod}, changing the distribution from
uniform to $a_0^{-1}$ per unit interval of $a_0$, by applying
appropriate weights to the integrated particles, changes $N_{JF}$ from
42 to 41 and $N_{HT}$ from 108 to 77 (cf.\ Table~\ref{dmod}).

Although our paper is mainly concerned with the origin of SP comets, we can
compare our results for various cometary populations with other
work. Fern\'andez et al.\ (2004), following Trujillo et al.\ (2000)
estimate 7.5$\times$10$^9$ objects with radius $R>1$ km and $q>30$ \AU\,
$a>50$ \AU\ but note an order of magnitude uncertainty in this number.
Moreover, our estimates are based on the flux of new comets with
$H_{10}<11$ corresponding to $R>0.3$ km according to Fernandez and Sosa
(2012). The number of such objects should be larger than the number
of objects with $R>1$ km.  Thus the estimate of Fern\'andez et al.\
(2004) does not contradict our possible values of $N_{\rm S}$. Estimates
for the number of comets in the outer Oort cloud range up to 10$^{12}$
(cf.\ Heisler 1990; Weissman 1996; Section 2.4 of Dones et al.\ 2004),
while the distribution of comets in different parts of the Oort cloud is
consistent with other models (cf.\ Emel'yanenko et al.\ 2007;
Dybczy\'nski et al.\ 2008; Leto et al.\ 2008).
 
Our data correspond to an initial population of approximately
1.6$\times$10$^{12}$ objects with $R>0.3$ km in the region
$25<q_0<36$ \AU, $50<a_0<300$ \AU. This is quite consistent with the
value of $\sim$\,3\,$\times$\,$10^{12}$ objects with $R>0.5$ km and
cometary albedos in the original trans-Neptunian planetesimal disc,
presented in Figure 1 of Morbidelli et al.\ (2009).

The data of Table~\ref{nmod} show that almost all JF comets originate
from orbits with initial perihelia in the outer planetary system, and
that over 90\% of the steady-state number of HT comets come from the
same $25<q_0<36$ \AU\ region.
This indicates that the majority of observed HT comets
would have had initial orbits with perihelion distances largely overlapping
the range of perihelia of the objects that eventually became JF comets.
This is in contrast to the general picture described in Section~1, where JF
comets largely originate from initial orbits in the trans-Neptunian region
and HT comets from initial orbits in the region of the giant planets, with
subsequent very different dynamical histories.

For all the models in Table \ref{dmod}, the orbital distributions of SP
objects with $q<1.5$ \AU\ coming from the Oort cloud have similar
characteristics.  Figures \ref{atall} and \ref{itall} show the orbital
element distributions in perihelia (i.e.\ equal weight to each perihelion
passage) for SP objects with $q<1.5$ \AU\ coming from the Oort cloud,
applying the restrictions $n_2 = 420$, $n_1 = 150$, $\beta=2$ (all objects
are equally presented, thus formally $\alpha=0$ in these plots).  The
Figures show that in our model, JF comets ($T>2$) are concentrated near the
ecliptic plane, approximately 70\% of them having $i<15^\circ$.
Regarding HT comets ($T<2$), although the model reveals both prograde
and retrograde orbits, prograde HT comets outnumber retrograde ones. In
these ways the basic features of these distributions are consistent with
those of the observed distributions in Figures
\ref{atql1.5} and \ref{itql1.5}.

In our model, all the modelled objects with periods under
20 yr have inclinations $i<60^\circ$. There are several reasons for
this. First, the majority of objects captured to the JF
population originate from the inner Oort cloud
(Emel'yanenko 2005). In our model, the inner Oort cloud is a rather
flattened source of comets (Emel'yanenko et
al.\ 2007). Secondly, the majority of such objects are
injected from the inner Oort cloud on to orbits with
perihelia in the region of the outer planets by external
perturbations. Their subsequent evolution is similar to the
scheme described for trans-Neptunian objects by
Kazimirchak-Polonskaya (1972) and Levison and
Duncan (1997). The latter showed that preferentially objects
with Tisserand parameters near 3 with respect to a planet
cross the orbit of this planet. This suggests that mainly
objects on prograde orbits are transferred to the inner
planetary region.

Our results -- from analysing observed SP comets -- about the initial
distribution of objects that form the Oort cloud are consistent with the
standard picture of the origin of the Solar system. The conclusion was
that $\beta \ge 1$: this corresponds to comets originally from the outer
planetary region having a greater probability of survival and thus a
longer lifetime as active comets, with objects originating from regions
with small heliocentric distances conversely becoming extinct more
quickly. This accords with the amount of water (as the main driver of
cometary activity) being larger for more distant objects in the early
Solar system.

\section{Summary and conclusions}

We have developed a model of the origin and evolution of the Oort cloud
which is consistent with the basic observed orbital distributions of
comets, Centaurs and high-eccentricity trans-Neptunian objects.
Rather than requiring intrinsically different fading properties for
Jupiter-family and Halley-type short-period comets, the model instead
adopts the hypothesis that the physical lifetime of objects as active
comets in the inner planetary region at the present epoch is a function
of their initial perihelion distance in the early Solar system, and is
the same for both JF and HT comets.  The observed JF and HT populations
also constrain the initial distribution of objects versus perihelion
distance.  Our results show that:

\begin{enumerate}

\item The mean physical lifetime of comets is $\ltsimeq$200 revolutions in
the region $q<1.5$ \AU. This implies a significant cometary contribution to
the distribution of small bodies (`boulders' and dust) making up the
near-Earth interplanetary complex.

\item No model in which the initial number of comets is a decreasing
function of their initial perihelion distance $q_0$ in the early Solar
system can explain the present observed distribution of short-period
comets.

\item Models in which the initial distribution of objects versus perihelion
distance is concentrated more towards the outer planetary region, and in
which their present active physical lifetime is an increasing function of
$q_0$, are consistent with the present orbital distributions and numbers of
both HT and JF comets.

\item Essentially all the observed HT comets and nearly half the observed
JF comets come from a proximate
Oort-cloud source (i.e.\ have experienced orbits
with $a > 10^3$ \AU).  The remaining $\sim$50\% of observed JF
comets come from the {\em observed\/} near-Neptune high-eccentricity
(NNHE) population, a dynamically unstable region in which the cometary
numbers decline by 95\% over 4 Gyr.  In addition, more than 90\%
of all Centaurs ($5\!<\!q\!<\!28$ \AU, $a<1000$ \AU) come from
the Oort cloud.

\item The model predicts that there is a significant Oort-cloud
contribution to the NNHE population.  The number of such objects is
comparable to the debiased number of objects already discovered in the
NNHE region, but they are still undetected owing to observational biases
(e.g.\ considering large semimajor axes or high inclinations).

\end{enumerate}

\begin{acknowledgements}
VVE would like to acknowledge the support provided by the Federal
Targeted Programme `Scientific and Educational Human resources of
Innovation-Driven Russia' for 2009--2013, and the STFC-funded visitor
programme of the Armagh Observatory.  Astronomy at Armagh Observatory is
supported by grant-in-aid from the Northern Ireland Department of Culture,
Arts and Leisure. We thank Julio Fern\'andez and Ramon Brasser for their
reviews.
\end{acknowledgements}



\end{document}